\def\eqnum#1{\eqno (#1)}
\def\fnote#1{\footnote}

\documentstyle{article}
\begin{document}

\noindent cond-mat/9212032 (to appear in ${\it J.}$ {\it Phys.: Cond. Matt})
\par
\medskip
\centerline{{\Large SPATIAL CORRELATION OF CONDUCTION ELECTRONS IN METAL WITH}
}
\medskip
\centerline{{\Large COMPLICATED GEOMETRY OF THE FERMI SURFACE.}
}
\medskip
\noindent D.I.Golosov\fnote{1}{\noindent Department  of  Physics,  Rutgers
University,  Piscataway, $NJ
08855-0849,$ U.S.A.
\par
\noindent Internet: golosov@physics.rutgers.edu
\par
}
 , M.I.Kaganov.
\par
\medskip
\noindent $P. L.$  Kapitza   Institute   for   Physical   Problems,   Russian
Academy of Sciences, Vorobyovskoye sh.  2, Moscow 117334 Russia
\par
\medskip
\centerline{{\it ABSTRACT}
}
\medskip
The  "density-density"  correlation  function  of  conduction
electrons  in  metal  is  investigated.  It  is  shown,  that  the
asymptotic behaviour of the $CF$ depends on the shape and the  local
geometry of the Fermi surface.  In  particular,  the  exponent  of
power law which describes the damping of Friedel  oscillations  at
${\it r \rightarrow \infty } (-{\it 4}$ for an isotropic Fermi  gas)  is
determined  by  local
geometry of the FS. The applications of the  obtained  results  to
calculations of the $CF$ in a metal near  the  electron  topological
transition and of the RKKY exchange  integral  are  considered  as
well.
\par
\medskip
\noindent {\it Classification numbers:} ${\it 71.45G, 71.25P.}$
\par
\medskip
\noindent {\bf 1. Introduction.}
\par
\medskip
In  this   paper   we   investigate   the   "density-density"
correlation function (CF) of conduction electrons in  a  metal  at
${\it T=0.}$
\par
It is well-known, that correlation function of  electron  gas
can be written as
\par
$$
\nu {\it (\vec{r}) \equiv } {{\it 1}\over <{\it n>}} < \Delta {\it
n(\vec{r}}_{1}{\it ) \Delta n(\vec{r}}_{2}{\it ) > - \delta (\vec{r}}_{1}{\it -
\vec{r}}_{2}{\it ) =}
$$
$$
{\it = -} {{\it 2}\over <{\it n>}} \mid \int {\it n}_{\vec{p}} \exp  {\it
(}\hbox{{\it i{p}{r}} ${\it / \hbar)} {{\it d}^{3} {\it p}\over {\it (2\pi
\hbar)}^{3}} \mid ^{2} {\it , \vec{r} = \vec{r}}_{2} {\it - \vec{r}}_{1} {\it
.}\qquad ^{\hbox{{\it (1)}}}$}
$$
\noindent Here the corner brackets stand for average, $\Delta {\it n(\vec{r}) =
n(\vec{r}) - <n>}$ is
the departure of electron density ${\it n(\vec{r})}$ from its average value
$<{\it n>}$,
\par
$$
{\it n}_{\vec{p}} {\it = \{} ^{1 {\it , \epsilon }(\vec{p}{\it ) < \epsilon
}_{F}}_{0{\it , \epsilon (}\vec{p}{\it ) > \epsilon }_{F}}
\eqnum{\hbox{{\it 2}}}$$
\noindent - Fermi distribution function, ${\it \vec{p}}$- momentum, $\epsilon
_{F}$  -  Fermi  energy.
Note, that equation ({\it 1}) is valid for any (not necessary isotropic)
dispersion law $\epsilon ({\it \vec{p}})$.
\par
Let us remind, that for $\epsilon {\it (\vec{p}) = p}^{2}{\it / 2m}$ {\it ,}
equation (1) leads to
the following expression for the $CF [1]$:
\par
$$
\nu {\it (\vec{r}) \approx } {{\it 3 \hbar}\over {\it 2\pi }^{2}{\it p}_{F}{\it
r}^{4}} \cos ^{2} {{\it p}_{F}{\it r}\over \hbar} {\it , r \gg  \hbar/p}_{F}
\eqnum{\hbox{{\it 3}}}$$
\noindent $({\it p}_{F}$ - Fermi momentum), which contains  Friedel
oscillations  (FO)
[2] with the wave number $2{\it p}_{F}/ \hbar$ .
\par
One can easily determine the  wave  numbers  of $FO ($if  the
direction of radius vector $\vec{r}$ is given) also  in  the  case  of  an
arbitrary dispersion law. Indeed, the long-wavelength behaviour of
the $CF ($equation  {\it (1)})  is  determined  by  the  singularities  of
the integral
\par
$$
{\it S(p}_{\parallel }{\it ) =} \int {\it n}_{\vec{p}} {\it d}^{2}{\it
p}_{\perp },\qquad {\it p}_{\parallel } {\it = \vec{p}\vec{r}/ r},\qquad {\it
\vec{p}}_{\perp } {\it = \vec{p} - p}_{\parallel }{\it \vec{r}/ r.}
\eqnum{\hbox{{\it 4}}}$$
\noindent as a function of ${\it p}_{\parallel }$ {\it .} Obviously, ${\it
S(p}_{\parallel }{\it )}$  is nothing but the  square
of the section of  the  Fermi  surface  (FS)  formed  by  a  plane
perpendicular to vector ${\it \vec{p}}$ and located at the distance ${\it
p}_{\parallel }$  from the
origin in ${\it \vec{p}}$-space. Therefore the singularities we  are
interested
in correspond to the tangencies of the $FS$ to planes  perpendicular
to ${\it \vec{r}}.$ The nature of  these  singularities  depends  on  the  {\sl
local}
geometry of the $FS$ near tangency points.
\par
Therefore the wave numbers of $FO \Delta {\it k}_{ij}$ are  the  distances  in
${\it \vec{p}}$-space between planes tangent to the $FS$ and  perpendicular  to
${\it \vec{r}}
($see  Fig.1).  It  is  well-known,  that  the  wave  numbers $\Delta {\it
k}_{ij}$
correspond also to Migdal - Kohn singularities [3] in the spectrum
of phonons propagating  in  the  direction ${\it \vec{r}/r}$,  as  well  as  to
singularities in  the  longitudinal  plasmon  spectrum  (cf.  [4])
etc\fnote{2}{\noindent Also, the quantities ${\it (\Delta k}_{ij}{\it )}
^{-1}$are proportional to the periods of
oscillations of the sound absorption  coefficient  in  a  magnetic
field [5].
\par
}
 . Thus, the vectors $\Delta {\it k}_{ij}{\it \vec{r}/r}$ define the
distinguished  points
in the {\it reciprocal} space; since the present paper is devoted to the
spatial correlation of electrons, we shall study the effects  that
occur in the {\it real} space.
\par
One can also come to the same conclusions  if  one  considers
the Fourier component of the $CF$
\par
$$
\nu {\it (\vec{k}) = -} {{\it 2}\over <{\it n>}} \int  {\it n}_{\vec{p}} {\it
n}_{\vec{p}+\hbar\vec{k}} {{\it d}^{3}{\it p}\over {\it (2\pi \hbar)}^{3}} {\it
,}
\eqnum{\hbox{{\it 5}}}$$
\noindent which is proportional to the volume of the intersection of the $FS$
and its analogue, shifted by the vector $\hbar{\it \vec{k}}.$ The singularities
 of
this quantity (as a function of ${\it \vec{k}} )$ correspond to  the
tangencies
of the $FS$ to its shifted analogue, as well as to ${\it \vec{k} = 0}.$  We
may
conclude tentatively, (see  section  2  below)  that  the  way  of
damping of $FO$ at ${\it r \rightarrow \infty }$ is determined by the local
geometry of the
$FS$ at the tangency points. Thus, the behaviour of the $CF$, as  well
as many  other  electronic  properties  of  metals  (cf.  [6])  is
determined by the $FS$ geometry.
\par
It turns out that one can describe the  asymptotic  behaviour
of the $CF$ at large ${\it r}$ with the formula
\par
$$
\nu {\it (\vec{r}) = -} {{\it 1}\over {\it 32\pi }^{4}} {{\it r}^{3}_{e}\over
{\it r}^{4}} \sum^{}_{ij} {\it A}^{*}_{i} {\it A}_{j} \exp  {\it (i \Delta
k}_{ij} {\it r)\quad .}
\eqnum{\hbox{{\it 6}}}$$
\noindent Here ${\it r}_{e} {\it = <n>}^{-1/3}$ is the average distance between
electrons,  and
indexes ${\it i}, {\it j}$ enumerate the tangency points. Factors ${\it A}_{i}
($which are,
in general, functions of ${\it r}$ as well as of ${\it \vec{r}/ r} )$  depend
on  the
local geometry of the $FS$ at the corresponding tangency points  and
have a dimensionality  of  inverse  length.  Due  to  the  central
symmetry  of  the $FS$,  expression   ({\it 6})   is   real   (obviously,
$\Delta {\it k}_{ij} {\it = - \Delta k}_{ji} )$.
\par
At ${\it i = j, \Delta k}_{ij} {\it = 0}$   and  corresponding  terms  in  the
$CF$
decrease monotoneously as ${\it r}$ increases. These terms are due to  the
singularity (namely, discontinuity of  the  first  derivative)  of
$\nu {\it (\vec{k})}$ at ${\it \vec{k} = 0.}$
\par
Usually, one  can  write  down  the  validity  condition  for
equation ({\it 6)} as
\par
$$
\Delta {\it k}_{ij}{\it r \gg  1} \hbox{for any ${\it i \neq  j}$  {\it .}}
$$
\noindent Note that the wave numbers $\Delta {\it k}_{ij} ($as well as  their
total  number)
depend on the direction of radius vector ${\it \vec{r}}$ .
\par
So far we discussed  the  electron  gas  with  an  arbitrary
dispersion law. The co-ordinate dependency of the $CF$ of conduction
electrons in a real metal is  not  exhausted  by  equation  ({\it 1}).In
order to describe this case, one  should  take  into  account  the
influence of periodic potential of crystal lattice not only on the
spectrum but also on the electron wave functions:  wave  functions
that should be used now are the  Bloch  waves  (instead  of  plane
waves ). This topic  is  considered  in  detail  in  our  previous
article [7].  The  qualitative  results  of  this  straightforward
though somewhat cumbersome consideration are listed below.
\par
1.Due to the  broken  translation  invariance,  the  co-ordinate
dependency of the $CF \nu {\it (\vec{r}}_{1} {\it , \vec{r}}_{2}{\it )}$ is not
reduced to  the  dependency
only on the difference of its arguments ${\it \vec{r} = \vec{r}}_{2}{\it -
\vec{r}}_{1}$  {\it .}  If  the
value  of ${\it \vec{r}}$ is fixed, $\nu ({\it \vec{r}}_{1} {\it
,\vec{r}}_{2}{\it )}$ is a periodic function of ${\it \vec{r}}_{1}$:
\par
$$
\nu {\it (\vec{r}}_{1}{\it ,\vec{r}}_{2}{\it ) = \nu (\vec{r}}_{1}{\it ,\vec{r}
+ \vec{r}}_{1}{\it ) = \nu (\vec{r}}_{1}{\it + \vec{a} , \vec{r} + \vec{r}}_{1}
{\it + \vec{a}) ,}
\eqnum{\hbox{{\it 7}}}$$
\noindent where {\it {a}} is any lattice period.
\par
2. Let us keep ${\it \vec{r}}${\it = const} and average $\nu {\it
(\vec{r}}_{1}{\it ,\vec{r}}_{2}{\it )}$ over ${\it \vec{r}}_{1}${\it .} Denote
the
obtained quantity  by $\nu ({\it \vec{r})}$.  Its  asymptotic  behaviour  will
be
similar to the one described by equation (6). Wave numbers  of $FO$
are to be determined in the same way - the only difference is that
now one should imagine  the $FS$  as  a  periodic  surface  in  the
reciprocal space (not  restricted  to  the  first  Brillouin  zone
alone!).  Therefore,  the  expression  for $\nu {\it (\vec{r})}$  should
contain
summation  over  reciprocal  lattice vector ${\it \vec{b}}$; terms with ${\it
\vec{b} \neq  0
(}${\it Umklapp} terms) oscillate with a wave number $\Delta {\it k}_{ij}{\it +
\vec{b}\vec{r}/ r.}$
\par
$3.CF$ contains  a  term,  whose  oscillation  wavelength  is  the
inverse diameter of the FS. Note, however, that if the $FS$ is open,
there can exist such a region of radius vector ${\it \vec{r}}$ directions,
where
$\nu ({\it \vec{r})}$ does not oscillate at all.
\par
Let us now consider the dependency of the $CF$  on  the  local
geometry of the FS. For the sake of simplicity, we shall  restrict
ourselves to the analysis of equations (1) and ({\it 6}); this  approach
may be easily generalized for the case of conduction electrons  in
a real metal [7].
\par
\medskip
\noindent {\bf 2. The Effect of Local Geometry of Fermi  Surface  on
Correlation
Function.}
\par
\medskip
{}From equation (6), it follows  that  each  term  in  the $CF$
contains the factors ${\it A}_{i}$ , which depend on the  local  geometry  of
the $FS$ at each  of  the  two  contributing  tangency  points.  The
expressions for ${\it A}_{i}{\it (\vec{r})}$ for various cases are listed
below. We shall
not describe here  the  calculations  -  one  might  find  such  a
description in [7].
\par
Let us keep the  direction  of  radius  vector ${\it \vec{r}}$  fixed  and
consider the factor ${\it A}_{i}$ , corresponding to the ${\it i}^{th}$
tangency point.
\par
(i) If it is an elliptic point, then
\par
$$
{\it A}_{\hbox{ell}}{\it = 2 G}^{-1/2}\quad ,
\eqnum{\hbox{{\it 8}}}$$
\noindent where ${\it G}$ is  the  Gaussian  curvature  (product  of  the
principal
curvatures [8]) of the FS.
\par
(ii) For a hyperbolic tangency point, one obtains
\par
$$
{\it A}_{\hbox{hyp}}{\it = \pm  2 i \mid G\mid }^{-1/2} {\it ,}
\eqnum{\hbox{{\it 9}}}$$
\noindent where the sign depends on the direction of radius vector  (whether
${\it \vec{r}}$ or $-{\it \vec{r}} )$.
\par
In these cases, {\it A} does not depend on ${\it r}$, and the amplitude of
$FO$ corresponding to a pair of elliptic and/or hyperbolic points of
tangency, decreases at ${\it r \rightarrow \infty }$ as ${\it r}^{-4}$.
\par
(iii) The domains of elliptic and hyperbolic points on the $FS$ are
separated by the  lines  of  {\sl parabolic}  points.  Up  to  very  few
exceptions $(Na, K, Cs, Rb$ and Bi),  the $FS$  of  real  metals  are
rather complex and do contain the lines of parabolic points [6,9].
\par
In the generic case, one can choose the local coordinates $\xi _{1} ,
\xi _{2}$ on the $FS$ in such a way, that the departure $\xi _{3}$ of the $FS$
from
the plane, tangent at parabolic point, reads:
\par
$$
\xi _{3}{\it (\xi }_{1} {\it , \xi }_{2} {\it ) \approx  C\xi }^{3}_{1} {\it -
B\xi }^{2}_{2} {\it .}
\eqnum{\hbox{{\it 10}}}$$
\noindent {\it (}here we restrict ourselves to the leading  order  in ${\it x},
{\it y})$.  We
shall assume, for convenience, that ${\it B > 0}$ and ${\it C > 0.}$
\par
If the tangency point is parabolic, then
\par
$$
{\it A}_{\hbox{par}}\approx  {{\it 2 {\cal K}(\sin  \pi /12) \Gamma (5/6)}\over
{\it 3}^{1/4} {\it C}^{1/3} {\it B}^{1/2} \pi } {\it r}^{1/6} \propto  {\it
r}^{1/6} {\it .}
\eqnum{\hbox{{\it 11}}}$$
\noindent Here ${\cal K}$  -  elliptic  integral, $\Gamma $   -   Euler's
gamma-function;
\par
$$
{\cal K}{\it (\sin  \pi /12) \Gamma (5/6) \approx  1.5 .}
$$
Thus we immediately notice, that if parabolic tangency points
appear at a given  direction  of ${\it \vec{r}}$  then  the  amplitude  of  the
corresponding term in the $CF$ decreases as ${\it r}^{-11/3}$, i.e. slower than
in the usual case. Due to the fact that parabolic points on the $FS$
are not isolated, but form continuous lines, there exist cones  of
radius  vector  directions  corresponding  to  the   presence   of
parabolic tangency points\fnote{3}{\noindent If the $FS$ contains parabolic
points then the  singularities  of
kinetic  characteristics  of  the  metal  are  enhanced  for   the
corresponding directions of the quasimomentum vector [10,11].
\par
}
 .
\par
Let us now denote by $\Delta \theta $ the angle between the  normal  at  the
parabolic point and the projection of radius vector onto the plane
perpendicular to the line of parabolic points (Fig.2). At $\Delta \theta  <
{\it 0,}$
there are no tangency points near the parabolic points. At $\Delta \theta  >
{\it 0,}$
however, there exist both elliptic and hyperbolic tangency  points
and both of them approach parabolic point as $\Delta \theta  \rightarrow {\it
0}.$ Therefore  in
the $CF$ there appears an additional long-wavelength term which  has
the wave number
\par
$$
{\it 2\cdot 3}^{-3/2} {\it C}^{{\it -}1/2} {\it (\Delta \theta )}^{3/2} .
$$
Due to the vicinity of a parabolic point, equations  ({\it 8})  and
({\it 9}) should be modified  (one  may  not  take  into  account  terms
quadratic  in ${\it x},{\it y}$ only).  Thus,  for  an  elliptic point
(Fig.$2,
{\it x < 0)}$, we have
\par
$$
{\it A}_{\hbox{ell}}\approx  {\it B}^{-1/2}(3C\Delta \theta )^{-1/4}{\it (1 -}
{{\it 5iC}^{1/2}\over {\it 16\cdot 3}^{1/2}\cdot \Delta \theta ^{3/2}} {{\it
1}\over {\it r}} {\it ) ,}
\eqnum{\hbox{{\it 12}}}$$
\noindent and for a hyperbolic point, we have
\par
$$
{\it A}_{\hbox{hyp}}\approx  {\it iB}^{-1/2}{\it (3C\Delta \theta )}^{-1/4}{\it
(1 +} {{\it C}^{1/2}\over {\it 2\pi 3}^{1/2}(\Delta \theta {\it )}^{3/2}} {\it
(1 -} {{\it 5\pi }\over {\it 8}} {\it i )} {{\it 1}\over {\it r}} {\it ) .}
\eqnum{\hbox{{\it 13}}}$$
\noindent This distinction between elliptic and hyperbolic points is not  so
surprising.  One  has  to   remember    that   the   singularities
of $\nu {\it (\vec{k})}$ corresponding to elliptic and hyperbolic  tangency
points
are somewhat different.
\par
Both formulae {\it (12)} and ({\it 13}) are valid at
\par
$$
\Delta \theta  \gg  {\it (r}^{2}{\it / C)}^{1/3}
\eqnum{\hbox{{\it 14}}}$$
Suppose that at $\Delta \theta  < {\it 0}$ there exist ${\it 2{\cal N}}$
tangency  points  (here
we are taking into account the symmetry of the $FS)$ and, therefore,
\par
$$
{\it 2{\cal N} - 1 +} {{\cal N}{\it ({\cal N} - 1)}\over {\it 2}} {\it =}
{{\cal N}{\it ({\cal N} + 3)}\over {\it 2}} {\it - 1}
$$
\noindent different wave numbers of FO. At $\Delta \theta  {\it = 0}$  there
appear   two  new
(parabolic) tangency points, and the amount  of $FO$  wave  numbers
increases by ${\cal N} + {\it 2}.$ Then at $\Delta \theta  > {\it 0}, \Delta
\theta  ^{<}_{\hbox{~}} {\it (r}^{2}{\it / C)}^{1/3}$ the  crossover
occurs (the contribution of each parabolic  tangency  point  turns
into the sum of contributions of elliptic and hyperbolic  tangency
points) and finally at $\Delta \theta  \gg  {\it (r}^{2}{\it / C)}^{1/3}$ there
are ${\it 2({\cal N} + 2)}$ tangency
points and\quad ${\it ({\cal N} + 2)({\cal N} + 5)/ 2 - 1}\quad FO$ wave
numbers.
\par
Let us consider the Gauss map of the $FS ($Fig. $3a)$. The  Gauss
map is a mapping of the surface onto the unit sphere,  induced  by
the normal at each point of the  surface  (the  direction  of  the
normal gives a point of the sphere, thus giving the  image  of  an
original point of the surface) [8]. The number of tangency  points
changes by 4 as the  point,  representing  the  direction  of  the
radius vector, crosses the line corresponding to the directions of
the normals at parabolic points. This is the  only  way  for  this
number to be changed. Suppose that for the direction of  a  radius
vector lying within the area 1 (see Fig. $3a)$ the total  number  of
the tangency points (on the  whole $FS)$  is $2{\cal N}$.  Then  for ${\it
\vec{r}/ r}$
situated on the lines BE or $ED$ there are $2({\cal N}{\it +1)}$  tangency
points
(two parabolic points have been added).  The  number  of  tangency
points for ${\it \vec{r}/r}$ lying in areas 2 and 4, on lines $AE$ or $EC$,
and  in
area 3 are given by ${\it 2({\cal N}+2)}, {\it 2({\cal N}+3)}$, and ${\it
2({\cal N}+4)}$ respectively.  The
angle $\theta $ appearing on Fig.2 is just the distance between point ${\it
\vec{r}/r}$
and the image of the line of parabolic points on the Gauss map.
\par
(iv) Points at which the $FS$ becomes flat  [12]  manifest  another
type of local geometry. The $FS$ with such a  point  is  a  boundary
case between convex surfaces and surfaces with $a '$crater'.  Near
this flattening point, the departure of the $FS$  from  the  tangent
plane is a fourth-order form of the local coordinates  within  the
surface. If the tangency point is a point of such a kind, then
\par
$$
{\it A}_{fl} \propto  {\it r}^{1/2} .
\eqnum{\hbox{{\it 15}}}$$
Let us assume for definiteness, that the departure of  the $FS$
from the plane tangent to the $FS$ at the flattening point, is given
in the leading order by
\par
$$
\xi _{3}{\it (\xi }_{1}{\it ,\xi }_{2}{\it ) \approx  D (\xi }^{2}_{1} {\it +
\xi }^{2}_{2}{\it ) = D \xi }^{2}_{\perp } .
\eqnum{\hbox{{\it 16}}}$$
\noindent Then as the (elliptic) tangency point  approaches  the  flattening
point, the Gaussian curvature decreases and
\par
${\it A}_{\hbox{ell}}\approx  {\it 2}^{5/6}{\it 3}^{-1/2}{\it D}^{-1/3} \Delta
\theta ^{-2/3}\quad {\it ,\qquad (17)}$
\par
\noindent Here $\Delta \theta $ is now the angle between the radius vector and
the  normal
to the $FS$ at the flattening point. Formula {\it (17)} is valid at
\par
$$
\Delta \theta  \gg  {\it D}^{1/4}{\it r}^{-3/4}\qquad {\it ,}
\eqnum{\hbox{{\it 18}}}$$
\noindent where the intermediate region is the region of  crossover  between
the dependencies {\it (15)} and {\it (17)} (see Fig. $3b)$.
\par
(v) Suppose that the lines of parabolic points on the $FS$  cross
at some point. In real metals such a crossing occurs rather often;
for example, the $FS$ of the form shown in Fig.4 is  encountered  in
$Mo, W$, paramagnetic $Cr [6,9,11]$.
\par
The departure of the $FS$ from the plane tangent to the $FS$  in
this point of crossing, is given by a third-order form and we have
\par
$$
{\it A}_{\hbox{or}} \propto  {\it r}^{1/3}\qquad {\it .}
\eqnum{\hbox{{\it 19}}}$$
\noindent In fact, this is a flattening point of another kind. The important
feature of flattening tangency points of any  type  is  that  they
correspond to isolated points on the Gauss map (see Fig.3).
\par
(vi) Let us assume that the $FS$ contains finite flat elements. If
the radius vector is parallel to the normal to such an element  of
the $FS$, it leads to the contribution of this  "tangency  area"  in
the $CF$:
\par
$$
{\it A}_{\hbox{plane}} \propto  {\it S}_{\hbox{plane}}{\it r} ,
\eqnum{\hbox{{\it 20}}}$$
\noindent where ${\it S}_{\hbox{plane}}$ is the square of this flat element.
\par
In fact, the $FS$ of such a kind can hardly be stable: in  this
case, Peierls transition  [13]  with  the  transformation  of  the
lattice structure should become energetically favourable (at least
at sufficiently low temperatures). It is easy to notice that there
should be some correlation between extremely slow damping of $FO$ as
${\it r \rightarrow \infty } ($here $\nu  \propto  {\it r}^{-2}\cos  {\it
(\Delta k}_{\hbox{plane}}{\it r)})$ and the possibility of  Peierls
transition.
\par
\medskip
\noindent At this point we finish our consideration of the local geometry of
the FS. Note, that we have arrived to the  remarkable  conclusion:
the exponent, which determines  the  damping  of $FO$  as ${\it r \rightarrow
\infty }${\it ,}
depends on the local geometry at the points of tangency. Thus, one
might build up a hierarchy of types of the $FS$ local geometry  with
respect to the long-distance behaviour of the $CF$:
\par
1.Spherical $FS$: Amplitude of $FO$ decreases as ${\it r}^{-4}$;  this
exponent
is given, generally, by elliptic or hyperbolic tangency points and
thus corresponds to a generic direction in $a {\it 3D}$ metal.
\par
2.Cylindrical $FS$: Amplitude of $FO$ decreases as ${\it r}^{-3}$. This  is
the
case of $a 2D$ metal, or the case  of  a  toroidal $FS$  cavity  (see
sect.5 of Ref. [7])\fnote{4}{\noindent Point at which the $FS$ becomes flat
also gives the ${\it r} ^{-3}$law  (see
Eqn (15)).
\par
}
 .
\par
3.Flat FS. Amplitude of $FO$ decreases as ${\it r}^{-2} ($one could imagine it
as $a FS$ of an $1D$ metal).
\par
\noindent From our consideration, it follows that  parabolic  points  (along
with crossing points of the lines of parabolic points)  should  be
placed "between" spherical and cylindrical FS.
\par
\medskip
\noindent {\bf 3. Some Applications.}
\par
\medskip
Let  us  consider  a  metal  near  the  electron  topological
transition (ETT) point [14]. The ETT is the restructuring  of  the
$FS$ which occurs when $\epsilon _{F} {\it = \epsilon }_{cr} (\epsilon _{cr}$ -
the value of electron  energy
that corresponds to Van Hove singularity).  Depending  on  whether
this singularity is due to an extremum or to a saddle point of the
electron  dispersion  law,  a  new  cavity  of  the\quad $FS$   appears
(disappears), or a "neck" connecting two $FS$  cavities  is  formed
(disrupted). The difference ${\it z = \epsilon }_{F} {\it - \epsilon }_{cr}$
depends  on  the  applied
pressure (or, for example, impurity concentration),  so  that  the
ETT can be realized experimentally (see Ref. [15]  and  references
there).   The   ETT   causes    distinctive    singularities    of
thermodynamical and kinetic quantities [14-16].
\par
Near the ETT point,  long-wavelength terms appear in  the $CF$
due to the existence of small diameters of  the  FS.  If  the  ETT
results in the appearance of a new cavity of the $FS$ at ${\it z > 0}$  ,
then
\par
$\nu _{\hbox{ETT}}{\it (\vec{r}) \approx  -} {{\it 4}\over <{\it n>\pi
}^{4}}\cdot {{\it z}\over \hbar^{2}}\cdot {{\it m}_{1}{\it m}_{2}{\it
m}_{3}\over {\it r}^{4} \sum^{i=3}_{i=1}{\it (m}_{i}\cos ^{2}\theta _{i}{\it
)}}\cos ^{2}\{{\it (2zr}\sum^{3}_{i=1}{\it m}_{i}\cos ^{2}\theta _{i}{\it
)}^{1/2}{\it / \hbar\} .
(21)}$
\par
\noindent Here ${\it m}_{1} {\it , m}_{2} {\it , m}_{3}$ are effective masses
and the axes ${\it \hat{x}}_{1} {\it ,\hat{x}}_{2} {\it , \hat{x}}_{3}
($where ${\it x}_{i} {\it = r \cos \theta }_{i} )$ are the main axes of the
tensor  of  inverse
effective masses, so that the electron  dispersion  law  near  the
point of extremum can be written as
\par
$$
{\it z \approx } {{\it (\Delta p}_{1}{\it )}^{2}\over {\it 2m}_{1}} {\it +}
{{\it (\Delta p}_{2}{\it )}^{2}\over {\it 2m}_{2}} {\it +} {{\it (\Delta
p}_{3}{\it )}^{2}\over {\it 2m}_{3}} {\it .}
\eqnum{\hbox{{\it 22}}}$$
\noindent If (instead of  an  extremum)  one  has  a  saddle  point  of  the
dispersion law, one should write instead of {\it (22):}
\par
$$
{\it z \approx } {{\it (\Delta p}_{1}{\it )}^{2} {\it + (\Delta p}_{2}{\it
)}^{2}\over {\it 2m}_{\perp }} {\it -} {{\it (\Delta p}_{3}{\it )}^{2}\over
{\it 2m}_{\parallel }} {\it +} {\beta \over {\it 4m}^{2}_{\parallel }}{\it
(\Delta p}_{3}{\it )}^{4}
\eqnum{\hbox{{\it 23}}}$$
\noindent (we assume that the $FS$ near the extremal point $\Delta {\it \vec{p}
= 0}$   possesses
the rotational symmetry with respect to the ${\it p}_{3}$ axis  and  is  also
symmetric with respect to the $\Delta {\it p}_{3} {\it = 0}$ plane). In  this
case,  the
"neck" of the $FS$, which exists at ${\it z > 0,}$ is  ruptured  at ${\it z < 0
(}$Fig.5). Therefore, at $\mid {\it z\mid  \ll  \epsilon }_{F}$ there exist
long-wavelength $(\Delta {\it k} ^{<}_{{\it\char"7E}}
{\it (2m\mid z\mid )}^{1/2}/ \hbar \ll  k_{F} {\it )}$ terms in the CF. The
issue  of  significance
here is that the angular dependence of these  terms  "rotates"  by
$\pi {\it /2}$ as the sign of ${\it z}$ changes.
\par
Indeed, for example, at ${\it z < 0}$ and ${\it \vec{r} \parallel
\vec{p}}_{3}$ , due to the  small
distance between the two sheets  of  the $FS$,  there  exist  long-wavelength
component   of    the\quad $FO$    with     corresponding
$\Delta {\it k = (2 m}_{\parallel }\mid {\it z\mid )}^{1/2}{\it / \hbar}$  and
amplitude
\par
$$
{\it m}^{2}_{\perp }\mid {\it z\mid r}^{-4}{\it / (2\hbar}^{2}\pi ^{4} <{\it n>
m}_{\parallel }{\it )}.
$$
\noindent There are  no  long-wavelength  oscillations at ${\it \vec{r} \perp
\vec{p}}_{3}$  .  Vice
versa, for ${\it z > 0,}$  the long-wavelength oscillations are absent  at
${\it \vec{r} \parallel  \vec{p}}_{3}$ . For ${\it z > 0}$  and ${\it \vec{r}
\perp  \vec{p}}_{3}$ , due to small diameter of  the $FS$
"neck", the following long-wavelength term in the $CF$ appears:
\par
$$
\nu _{\hbox{ETT}}{\it (r) \approx  -} {{\it m}_{\parallel } {\it z}\over \pi
^{4}<{\it n>r}^{4}\hbar^{2}} \sin ^{2}\pmatrix{{\it r}&{\it (2m}_{\perp }{\it
z)}^{1/2}} \hbox{at ${\it \vec{r} \perp  \vec{p}}_{3}$ {\it .}}
$$
At $\beta {\it z > 0}$ {\it ,} the fourth-order term in ({\it 23)} gives rise
to  the
appearance of the lines of parabolic points on the $FS$  near ${\it
\vec{p}=\vec{p}}_{cr}
[16]$\fnote{5}{\noindent Note that  parabolic  points  on  the $FS$  themselves
 are  not
peculiar from the  point  of  view  of  electron  dispersion  law.
Generally, the lines of parabolic points may have  nothing  to  do
with ETT or Van Hove singularities (see sect.2 above).
\par
}
 (see Fig.5). Thus for some directions  of ${\it \vec{r}}$  forming  angle
$\theta _{\hbox{par}}$ with ${\it p}_{3}$ axis, the amplitude of the
long-wavelength  term  in
the $CF$  behaves  as ${\it r}^{-11/3}{\it = r}^{-4}{\it r}^{2/3}$  instead  of
${\it r}^{-4}$.  In  the
neighbourhood  of  these  peculiar  directions,  there  exists  an
additional small wave number of $FO$, which is due  to  neighbouring
tangency points pairs (elliptic and hyperbolic). At $\theta {\it =\theta
}_{\hbox{par}}$ ,  each
pair "sticks together", giving the parabolic  tangency  point  and
then disappearing.
\par
On the whole, the ETT, being a local  "event"  in  reciprocal
space, in co-ordinate space results in dramatic  restructuring  of
the long-wavelength terms in the CF.
\par
Obviously, the  knowledge  of  the $CF$  is  useful  when  one
considers  various  collective  phenomena  in  metals:  screening,
structure of spin glasses etc. Let us mention for example the RKKY
interaction [17-19]. This indirect  exchange  interaction  between
the site (or impurity) spins is  caused  by  exchange  interaction
between the localized spins and conduction electrons:
\par
$$
{\cal H} {\it = (J/ <n>)} \sum^{}_{i} \vec{\sigma } {\it \vec{S}}_{i} \delta
{\it (\vec{r} - \vec{r}}_{i}{\it ) .}
\eqnum{\hbox{{\it 24}}}$$
\noindent Here ${\it J}$ - exchange constant, $\vec{\sigma }$ - Pauli matrices,
${\it \vec{S}}_{i}$ - spin  of  the
ion located in the site ${\it \vec{r}}_{i}$. The RKKY exchange integral reads
\par
\noindent ${\it J}_{\hbox{RKKY}}{\it (\vec{r})= -2(}{{\it J}\over <{\it
n>}}{\it )}^{2} \int  {{\it n}_{\vec{q}} {\it (1-n} _{\vec{q}'}{\it )}\over
\epsilon {\it (\vec{q}) - \epsilon (\vec{q}')}} \exp {\it
[i(\vec{q}-\vec{q}')\vec{r}/ \hbar]} {{\it d}^{3} {\it q d}^{3} {\it q'}\over
{\it (2\pi \hbar)}^{6}}{\it ,
(25)}$
\par
\noindent and its Fourier component,
\par
\noindent ${\it J}_{\hbox{RKKY}}{\it (\vec{k})= -2(}{{\it J}\over <{\it
n>}}{\it )}^{2} \int  {{\it n}_{\vec{q}}{\it (1-n}_{\vec{q}-\hbar\vec{k}}{\it
)}\over \epsilon {\it (\vec{q})-\epsilon (\vec{q}-\hbar\vec{k})}} {{\it
d}^{3}{\it q}\over {\it (2\pi \hbar)}^{6}}{\it .\qquad (26)}$
\par
\noindent This kind of integral appears also in studying  of  the
electron-phonon interaction. It has singularities at  the  same  points  as
does the $CF ($except ${\it k = 0)}$. These singularities are called  Migdal
- Kohn singularities [3] if the electron velocity vectors  in  the
two contributing tangency points are antiparallel.  On  the  other
hand, if they are parallel, then the singularity is called  Taylor
singularity [20] and these two cases should be treated separately.
\par
It turns out, that at least for the great  majority  of  cases
there exists a quite simple relationship between  the  asymptotics
of $\nu {\it (\vec{r})}$ and ${\it J}_{\hbox{RKKY}}{\it (\vec{r})}${\it .}
Indeed, for a fixed direction of the  radius
vector
\par
$$
{\it J}^{(\Delta k)}_{\hbox{RKKY}} \propto  {\it r\nu }^{(\Delta k)}{\it
(r),\quad \Delta k \neq  0}
\eqnum{\hbox{{\it 27}}}$$
\noindent where ${\it J}^{(\Delta k)}_{\hbox{RKKY}}$ and $\nu ^{(\Delta k)}{\it
(r)}$ stand for the terms in ${\it J}_{\hbox{RKKY}}$ and  the $CF$
respectively, oscillating with the wave number $\Delta {\it k}.$
\par
It is well-known, that magnetic order  induced  by  the  RKKY
interaction (namely, the helicity vector) depends on the  geometry
of the $FS [18]$. Thus,  the  helicity  vector  is  expected  to  be
determined by the small diameter of the $FS [21]$.  Therefore  the
knowledge of the $FS$ and the $CF$ seems to allow  one  to  make  some
predictions  about  the  magnetic  arrangement.  We   expect,   in
particular, that  the  ETT  may  manifest  itself  in  changes  of
magnetic ordering.
\par
\medskip
\noindent {\bf 4.Conclusion.}
\par
\medskip
In this paper, we  neglected  temperature  effects,  electron
scattering  and  electron-electron  interaction.  The  first   two
effects lead to the  appearance  of  an  exponentially  decreasing
factor  in  the $CF$  as ${\it r\rightarrow\infty }$.   The   third   (i.e.
Fermi-liquid
interaction), probably leads  to   the    renormalization  of  the
coefficients (see, for example, [22]).
\par
Anyway, the dependence of the  asymptotic  behaviour  of  the
correlation function on  the $FS$  geometry  (including  the  local
geometry) survives after taking these effects into  account.  This
dependence should affect collective phenomena in  metals.  Indeed,
one has  to  know  the $CF$  to  construct  theories  of  alloying,
screening, exchange  magnetism  etc.  We  have  investigated  some
features of the $CF$ that seem to be interesting from this point  of
view. Among them are:
\par
1.  The  existence  of  the   cones   of   peculiar   directions,
corresponding to the lines of parabolic  points  on  the  FS.  For
these directions of the radius-vector, the $CF$ shows some  specific
features:
\par
(i) the attenuation of $FO$ is determined by a  factor  of ${\it r}^{-11/3}$
instead of usual ${\it r}^{-4}$; we doubt if this small  difference  has  any
observable effect.
\par
(ii) the number of $FO$ periods changes as the  direction  of  the
radius vector crosses these cones.
\par
(iii)On one side of such a cone, one of $FO$  periods  appears  to
become large (it goes to infinity as the direction of ${\it \vec{r}}$
approaches
the cone).
\par
2. The existence  of  the  isolated  peculiar  directions.  These
correspond to the crossings of the lines of  parabolic  points  on
the $FS ($flattening points).
\par
3. The restructuring  of  the $CF$  due  to  electron  topological
transition.
\par
\noindent We have also provided one example showing how an asymptote of  the
RKKY exchange integral depends on the asymptotic behaviour of  the
CF.
\par
It is a pleasure to thank M.Yu.Kagan  and  L.P.Pitaevsky  for
fruitful discussions. We are  also  grateful  to $A.$  Prakash  for
assistance in preparing this manuscript for publication.
\par
\medskip
\centerline{{\bf References}
}
\medskip
\noindent [1]  Landau L.D. and Lifshitz $E. M. 1969$ {\it Statistical Physics
(Mass.:
Addison-Wesley)}
\par
\noindent [2]  Friedel $J.1952$ {\it Phil. Mag.:} {\sl 43} 153
\par
1954 {\it Phil. Mag. Suppl.:} {\sl 3} 446
\par
1958 {\it Nuovo Cimento Suppl.:} {\sl 7} 287
\par
\noindent [3]  Migdal $A. B. 1958$ {\it Sov. Phys. JETP :} {\sl 7} 996
\par
Kohn $W. 1959$ {\it Phys. Rev. Lett.:} {\sl 2} 393
\par
\noindent [4]  Lifshitz $E. M.$  and  Pitaevski $L. P. 1981$  {\it Physical
Kinetics
(Oxford, New York: Pergamon)}
\par
\noindent [5]  Gurevich $V. L. 1960$ {\it Sov. Phys. JETP:} {\sl 10} 51
\par
\noindent [6]  Lifshits $I. M.$ Azbel' $M.$ Ya. and Kaganov $M. I. 1973$ {\it
Electron  theory
of Metals (New York: Consultants Bureau)}
\par
\noindent [7]  Golosov $D. I.$ and Kaganov $M. I. 1992$  {\it Sov.  Phys.
JETP:} {\sl 74} 186
\par
\noindent [8]  Dubrovin $B. A.$ Fomenko $A. T.$ and Novikov $S. P. 1992$ {\it
Modern Geometry
- Methods and Applications (New York: Springer)}
\par
\noindent [9]  Cracknell $A. P.$ and Wong $K. C. 1973$ {\it The  Fermi  Surface
 (Oxford:
Clarendon)}
\par
\noindent [10] Avanesyan G.T. Kaganov $M. I.$ and Lisovskaya $T.$ Yu. 1978 {\it
Sov. Phys.
JETP:} {\sl 48} 900
\par
\noindent [11] Kaganov $M. I.$  Kontorovich $V. M.$  Lisovskaya $T.$  Yu.  and
Stepanov $N. A. 1983$ {\it Sov. Phys. JETP:} {\sl 58} 975
\par
\noindent [12] Kaganov $M. I.$ and Lifshits $I. M. 1979$ {\it Sov. Phys. Usp.:}
{\sl 22} 904
\par
\noindent [13] Peierls $R. E. 1955$ {\it Quantum Theory of Solids (Oxford:
Clarendon)}
\par
\noindent [14] Lifshitz $I. M. 1960$ {\it Sov. Phys. JETP:} {\sl 11} 1130
\par
\noindent [15] Varlamov $V. A.$ Egorov $V. S.$ and Pantsulaya $A. V. 1989$ {\it
Adv. Phys.} {\sl 38}
469
\par
\noindent [16] Kaganov $M. I.$ and Gribkova Yu. $V. 1991$ {\it Sov} ${\it J}$
{\it Low Temp Phys} {\sl 17} 473
\par
\noindent [17] Ruderman $M. A.$ and Kittel $C. 1954$ {\it Phys. Rev.:} {\sl 96}
99
\par
Kasuya $T. 1956$ {\it Progr. Theor. Phys.:} {\sl 16} 45
\par
Yosida $K. 1957$ {\it Phys. Rev.:} {\sl 106} 893
\par
\noindent [18] Yosida $K. 1964$  {\it in:  Progress  in  Low  Temperature
Physics
(Amsterdam: North-Holland) ed.} ${\it C}$ {\it Corter} Vol. $4 p. 265$
\par
\noindent [19] Abrikosov $A. A. 1988$  {\it Fundamentals  of  the  Theory  of
Metals}
({\it Amsterdam - New York: North Holland)}
\par
\noindent [20] Taylor $P. L. 1963$ {\it Phys. Rev.:} {\sl 131} 1995
\par
Kaganov $M. I.$ Plyavenek $A. G.$ and Hitschold $M. 1982$  {\it Sov.  Phys.
JETP:} {\sl 55} 1167
\par
\noindent [21] Dzyaloshinskii $I. E. 1965$ {\it Sov. Phys. JETP:} {\sl 20} 665
\par
\noindent [22] Kaganov $M. I.$ and $M$\"o bius $A. 1984$ {\it Sov. Phys. JETP:}
{\sl 59} 405
\par
\medskip
\centerline{{\bf Figure Captions}
}
\medskip
\noindent {\sl Fig.1} Rather simple $FS ($"dumb-bell") for the  radius  vector
${\it \vec{r}\parallel }\vec{r}_{1}$
generates 8 different wave numbers of Friedel oscillations (due to
the symmetry,
\par
$$
\Delta {\it k}_{12} {\it = \Delta k}_{56} , \Delta {\it k}_{13} {\it = \Delta
k}_{46} , \Delta {\it k}_{14} {\it = \Delta k}_{25} {\it = \Delta k}_{36} ,
$$
$$
\Delta {\it k}_{15} {\it = \Delta k}_{26}, \Delta {\it k}_{23} {\it = \Delta
k}_{45} \hbox{and $\Delta {\it k}_{24} {\it = \Delta k}_{35} )$.}
$$
\noindent Circled numbers enumerate the tangency points for this  case.  The
total number of wave numbers depends on the direction  of ${\it \vec{r}}$:  for
${\it \vec{r}\parallel \vec{r}}_{2}$ there is the only wave number $\Delta {\it
k}.$
\par
\medskip
\noindent {\sl Fig.2.} Parabolic point on the $FS ($placed at the  origin  of
local
co-ordinates $\xi _{1} , \xi _{2} , \xi _{3}$ , where $\xi _{3}$ is parallel to
 the  normal).
For a small value of angle $\theta  > {\it 0},$ there are two tangency points
in
the vicinity of the parabolic point: elliptic (with $\xi _{1} < {\it 0})$  and
hyperbolic (with $\xi _{1} > {\it 0})$, and,  correspondingly,  the  small
wave
number $\Delta {\it k \propto  \theta }^{3/2}$. As $\theta  \rightarrow {\it
+0},$ these tangency points  approach  each
other; at $\theta  {\it = 0}$ they coincide at the parabolic point; at $\theta
< {\it 0},$
there are no tangency points in the  vicinity  of  this  parabolic
point (see also Fig. $3a)$.
\par
\medskip
\noindent {\sl Fig.3.} Gauss maps of the elements of the FS.
\par
${\sl 3a.}$ The image of  the  part  of  the $FS$  with  two  lines  of
parabolic points $AC$ and $BD$ intersecting in the point $E.$ Each point
in areas 2 and 4 is the image of the two points from the  original
part of the $FS$, each point in the area  3  -  the  image  of  four
points and each point in area 1 does not have any original  there.
Shaded areas correspond  to  crossover  (the  inequality  {\it (14)}  is
broken there). This obviously is only one  of  the  two  kinds  of
intersections of the lines of parabolic points.
\par
${\sl 3b.}$ Gauss map of the part of the $FS$ containing  the  point  at
whichthe $FS$  becomes  flat.  The  number  of  tangency  points  is
constant for ${\it \vec{r}/r}$ laying in this region of  directions.  The
shaded
area corresponds to crossover (inequality {\it (18)} is broken there).
\par
\medskip
\noindent {\sl Fig.4} Crossing of the lines of parabolic points on the $FS
[11]$.
\par
${\sl 4a.}$ Fragment of the $FS ($"octahedron") in a metal belonging  to
the molybdenum group.
\par
${\sl 4b.}$ Crossing of lines of parabolic points on  this  cavity  of
the $FS ($view from above).
\par
\medskip
\noindent {\sl Fig.5} Disappearance  ({\sl a})  and  appearance  ({\sl b})  of
the  lines  of
parabolic points (marked by the thick  lines)  during  the  "neck"
formation [16].
\par
\end{document}